\documentclass[12pt]{article}
\usepackage{tikz}
\usepackage{epsf}
\usepackage{amsmath}
\def\hybrid{\topmargin 0pt      \oddsidemargin 0pt
	\headheight 0pt \headsep 0pt
	\textheight 9in         
	\textwidth 6.25in       
	\marginparwidth .875in
	\parskip 5pt plus 1pt   \jot = 1.5ex}

\catcode`\@=11
\def\marginnote#1{}
\newcount\hour
\newcount\minute
\newtoks\amorpm
\hour=\time\divide\hour by60
\minute=\time{\multiply\hour by60 \global\advance\minute by-\hour}
\edef\standardtime{{\ifnum\hour<12 \global\amorpm={am}%
	\else\global\amorpm={pm}\advance\hour by-12 \fi
	\ifnum\hour=0 \hour=12 \fi
	\number\hour:\ifnum\minute<10 0\fi\number\minute\the\amorpm}}
\edef\militarytime{\number\hour:\ifnum\minute<10 0\fi\number\minute}

\def\draftlabel#1{{\@bsphack\if@filesw {\let\thepage\relax
   \xdef\@gtempa{\write\@auxout{\string
      \newlabel{#1}{{\@currentlabel}{\thepage}}}}}\@gtempa
   \if@nobreak \ifvmode\nobreak\fi\fi\fi\@esphack}
	\gdef\@eqnlabel{#1}}
\def\@eqnlabel{}
\def\@vacuum{}
\def\draftmarginnote#1{\marginpar{\raggedright\scriptsize\tt#1}}

\def\draft{\oddsidemargin -.5truein
	\def\@oddfoot{\sl preliminary draft \hfil
	\rm\thepage\hfil\sl\today\quad\militarytime}
	\let\@evenfoot\@oddfoot \overfullrule 3pt
	\let\label=\draftlabel
\let\marginnote=\draftmarginnote
	\let\marginnote=\draftmarginnote
   \def\@eqnnum{(\theequation)\rlap{\kern\marginparsep\tt\@eqnlabel}%
\global\let\@eqnlabel\@vacuum}  }

\def\numberbysection{\@addtoreset{equation}{section}
\def\theequation{\thesection.\arabic{equation}}}

\def\underline#1{\relax\ifmmode\@@underline#1\else
	$\@@underline{\hbox{#1}}$\relax\fi}

\def\titlepage{\@restonecolfalse\if@twocolumn\@restonecoltrue\onecolumn
     \else \newpage \fi \thispagestyle{empty}\c@page\z@
	\def\thefootnote{\fnsymbol{footnote}} }

\def\endtitlepage{\if@restonecol\twocolumn \else  \fi
	\def\thefootnote{\arabic{footnote}}
	\setcounter{footnote}{0}}  
\catcode`@=12
\relax

\def\beq{\begin{equation}}
\def\eeq{\end{equation}}
\def\bea{\begin{eqnarray}}
\def\eea{\end{eqnarray}}
\def\nn{\nonumber}

\relax
\hybrid

\begin{document}

\begin{titlepage}
\begin{center}
April~2023 \hfill . \\[.5in]
{\large\bf Correlation function for the punctual state of the fermion string in the space of dimension $D=10$. }
\\[.5in] 
{\bf Vladimir S.~Dotsenko}\\[.2in]
{\it LPTHE, 
Sorbonne Universit{\'e}, CNRS, UMR 7589\\
4 place Jussieu,75252 Paris Cedex 05, France.}\\[.2in]
               
 \end{center}
 
\underline{Abstract.}

Correlation function is defined and calculated for the punctual states of the fermion
supersymmetric string $(N=1)$, in its critical dimension $D=10$.

\end{titlepage}

\newpage

\numberwithin{equation}{section}

\section{Introduction.}

Usually, for the strings, one studies the scattering amplitudes of physical onshell states. 
On the contrary, in this paper we want to adresse the question of the correlation functions for strings, in Euclidian space, of critical dimensions $D=26$ for the bosonic string and $D=10$ for the fermionic string. And this is for a particular case of the two-point correlation function of punctual states.  

For the bosonic string, in dimension $D=26$, this function has already been considered in the literature long ago [1]. It is known to be pathological, divergent.

We want to study a similar two point function, but for the fermionic string in its critical dimension $D=10$.

But to start with, we shall reconsider in the next Section, with our technique, the bosonic string two point function for the punctual states. This is with the fully covariant formulation, with the functional integral for the $X^{\mu}(t,x)$ string position fields, where all $\mu$ components are integrated over, plus the ghost fields $b(t,x),\bar{b}(t,x),c(t,x),\bar{c}(t,x)$ which are also integrated over, which are propagating in a way that they are compensating the two 
nonphysical modes of the $X^{\mu}$ fields.

In the Section 3 we shell develop a similar technique for the fermion string two-point function, with $X^{\mu}$ and $\psi^{\mu}$ fields plus the corresponding ghosts.
We shall define and calculate the two-point correlation function for the punctual states of the fermion string.
The correlation function we obtain is not divergent, is well defined.

The Section 4 is devoted to discussions and conclusions.

 \section{Bosonic string two-point function}
 
 The field $X^{\mu}(t,x)$ and the ghosts $b(t,x),\bar{b}(t,x),c(t,x),\bar{c}(t,x)$, are defined on the cylinder -- on the strip $0<x<1$, $0<t<T$, Fig.1, 
 all the fields being periodic functions on $x$:
 \bea
 X^{\mu}(t,1)=X^{\mu}(t,0),\quad b(t,1)=b(t,0)\nn\\
 c(t,1)=c(t,0),\quad\bar{b}(t,1)=\bar{b}(t,0),\quad\bar{c}(t,1)=\bar{c}(t,0).
 \eea
 
 On $t$, the field $X^{\mu}(t,x)$  has fixed boundary conditions:
 \beq
 X^{\mu}(0,x)=0^{\mu}, \quad X^{\mu}(T,x)=R^{\mu}
 \eeq
 which corresponds to the string being punched at points $0^{\mu}$ and $R^{\mu}$ in the $X^{\mu}$ space, Fig.2.

\begin{figure}
\begin{center}
\begin{tikzpicture}[baseline=(current  bounding  box.center),  thick, scale = .6]
\draw (-1,6) node {$T$};
\draw (0.6,8.3) node {$t$};
\draw (-0.7,-0.7) node {$0$};
\draw (3,-0.7) node {$1$};
\draw (6.3,0.8) node {$x$};
\draw (-1,0)  -- (6,0);
\draw (0,6)  -- (3,6);
\draw (0,-1)  -- (0,8);
\draw (3,0) -- (3,6);
\draw [-stealth](4,0) -- (6.1,0);
\draw [-stealth](0,6) -- (0,8.1);
\draw[black,fill=black] ( 0,0) circle (0.5ex);
\draw[black,fill=black] ( 0,6) circle (0.5ex);
\draw[black,fill=black] ( 3,0) circle (0.5ex);
\end{tikzpicture}
\end{center}
\caption{The strip $0 < x < 1, 0< t < T$ in the parameter space of the variables $x,t$.}
\label{Fig1}
\end{figure}
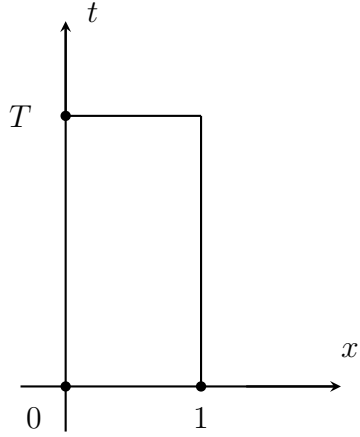

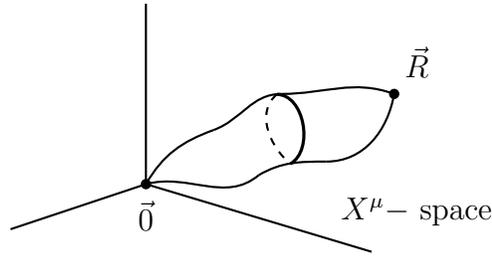
\begin{figure}
\begin{center}
\begin{tikzpicture}[baseline=(current  bounding  box.center),  thick, scale = .6]
\draw (0,0)  -- (0,4);
\draw (0,0)  -- (-3,-1);
\draw (0,0)  -- (5,-1.5);
\draw (0.,-0.7) node {$\vec{0}$};
\draw (6.,2.7) node {$\vec{R}$};
\draw (6.,-0.6) node {$X^\mu-$ space};
\draw[black,fill=black] ( 0,0) circle (0.5ex);
\draw[black,fill=black] ( 5.5,2) circle (0.5ex);
\draw[thick] (0,0) to [out=15,in=220] (2.5,0.2);
\draw[thick] (2.5,0.2) to [out=30,in=180] (4.0,0.5);
\draw[thick] (4.0,0.5) to [out=0,in=260] (5.5,2);
\draw[thick] (0,0) to [out=60,in=200] (1.5,1.2);
\draw[thick] (1.5,1.2) to [out=20,in=180] (3.0,2.);
\draw[thick] (3.0,2.0) to [out=0,in=160] (5.5,2);
\draw[very thick] (2.9,1.99) to [out=0,in=30] (3.2,0.45);
\draw[dashed] (2.9,1.99) to [out=220,in=140] (3.2,0.45);
\end{tikzpicture}
\end{center}
\caption{The punctual states correlator $<p(\vec{R}) p(\vec{0})>$ of the bosonic string, eq.~2.3.}
\label{Fig2}
\end{figure}

 We are looking for the correlation function which we define as
 \bea
 <p(R^{\mu})\cdot p(0^{\mu})>=\int^{\infty}_{0}dTe^{{-A}[X_{cl}]} G_{X}(R;T)\cdot G_{b,c}(R;T)\nn\\
 =\int_{0}^{\infty}dT\int DX^{\mu}(t,x)\int Db(t,x)\int D\bar{b}(t,x)\int Dc(t,x)\int D\bar{c}(t,x)\nn\\
 \exp\{-A[X^{\mu}]-B[b,\bar{b},c,\bar{c}]\}
 \eea
 with the actions:
 \beq
 A[X^{\mu}]=\int^{T}_{0}dt\int^{1}_{0}dx((\partial_{t}X)^{2}+(\partial_{x}X)^{2})
 \eeq
 \beq
 B[b,\bar{b},c,\bar{c}]=\int^{T}_{0}dt\int^{1}_{0}dx(b\bar{\partial}c+\bar{b}\partial\bar{c})
 \eeq
 
 \vskip1cm
 
 {\bf 1.}
 
 We calculate first the $X^{\mu}$ part of the correlator (2.3), 
 with the propogator, when $T$ is fixed,
 \beq
 G_{X}(\vec{R};T)=\int DX^{\mu}(t,x)\exp\{-A[X^{\mu}]\}
 \eeq
 We take
 \beq
 X^{\mu}(x,t)=X_{cl}(x,t)+X^{\mu}_{qu}(x,t)
 \eeq
 \beq
 X^{\mu}_{cl}(x,t)=(  X^{\mu}_{(2)}  -  X^{\mu}_{(1)})\frac{t}{T}+X^{\mu}_{(1)}
 \eeq
 with
 \beq
 X^{\mu}_{(1)}=0^{\mu},  \quad   X^{\mu}_{(2)}=R^{\mu}
 \eeq
 \beq
 X^{\mu}_{qu}(0,x)=X^{\mu}_{qu}(T,x)=0^{\mu}
 \eeq
 We have, evidently, for  $A[X^{\mu}]$ in (2.4), the classical part
 \beq
 A[X^{\mu}_{cl}]=T\cdot\frac{1}{T^{2}}(X^{\mu}_{(2)}-X^{\mu}_{(1)})^{2}
 =\frac{1}{T}(R^{\mu})^{2}
 \eeq
 \beq
 \exp(-A[X^{\mu}_{cl}])=\exp(-\frac{1}{T}(R^{\mu})^{2})
 \eeq
 
 For the quantum part, with $X^{\mu}_{qu}(t,x)$ being periodic function of $x$,
 it could be decomposed as follows:
 
 \bea
 X^{\mu}_{qu}(t,x)=\sum^{\infty}_{n=1} a^{\mu}_{n}(t)\sqrt{2}\sin(2\pi nx)
  + \sum^{\infty}_{n=1}b^{\mu}_{n}(t)\sqrt{2}\cos(2\pi nx) + b^{\mu}_{0}(t)
 \eea
 
 One obtains:
 \bea
 \partial_{x} X^{\mu}_{qu}(t,x)=\sum^{\infty}_{n=1}a^{\mu}_{n}(t)\sqrt{2}\cos(2\pi nx)
 \cdot 2\pi n \nn\\ -\sum^{\infty}_{n=1}b^{\mu}_{n}(t)\sqrt{2}\sin(2\pi nx)\cdot 2\pi n
 \eea

 \bea
 \partial_{t} X^{\mu}_{qu}(t,x)=\sum^{\infty}_{n=1} \partial_{t} a^{\mu}_{n}(t)\sqrt{2}\sin(2\pi nx)
 \nn\\ +\sum^{\infty}_{n=1} \partial_{t} b^{\mu}_{n}(t)\sqrt{2}\cos(2\pi nx) 
 + \partial_{t} b^{\mu}_{0}(t)
 \eea

Integrating over $x$, one obtains:

 \beq
 \int^{1}_{0}dx(\partial_{x}X^{\mu}_{qu}(t,x))^{2}=\sum^{\infty}_{n=1}4\pi^{2}n^{2}[(a^{\mu}_{n}(t))^{2}+(b^{\mu}_{n}(t))^{2}]
 \eeq
 
 \beq
 \int^{1}_{0}dx(\partial_{t}X^{\mu}_{qu}(t,x))^{2}=\sum^{\infty}_{n=1}[(\partial_{t}a^{\mu}_{n}(t))^{2}+(\partial_{t}b_{n}^{\mu}(t))^{2}] + (\partial_{t}b_{0}^{\mu}(t))^{2}
 \eeq
 
 \bea
 A[X^{\mu}_{qu}]=\int^{T}_{0}dt\int^{1}_{0}dx[(\partial_{x}X^{\mu}_{qu})^{2}+(\partial_{t}X^{\mu}_{qu})^{2}]\nn\\
 =\int^{T}_{0}dt\sum^{\infty}_{n=1}[4\pi^{2}n^{2}(a^{\mu}_{n}(t))^{2}
 +(\partial_{t}a^{\mu}_{n}(t))^{2}]\nn\\
 +\int^{T}_{0}dt\sum^{\infty}_{n=1}[4\pi n^{2}(b^{\mu}_{n}(t))^{2}+(\partial_{t}b^{\mu}_{n}(t))^{2}]+\int^{T}_{0}dt(\partial_{t}b^{\mu}_{0}(t))^{2}
 \eea
 With the boundary conditions zero, at $t=0$ and at $t=T$, the coefficients
  $a^{\mu}_{n}(t)$, $b^{\mu}_{n}(t)$, $b^{\mu}_{0}(t)$, in $X^{\mu}_{qu}(t,x)$, eq.(2.13),   
  can be decomposed on sinus:
 \bea
 a^{\mu}_{n}(t)=\sum^{\infty}_{m=1}a^{\mu}_{nm}\sqrt{\frac{2}{T}}\sin(\frac{\pi mt}{T})\nn\\
 b^{\mu}_{n}(t)=\sum^{\infty}_{m=1}b^{\mu}_{nm}\sqrt{\frac{2}{T}}\sin(\frac{\pi mt}{T})\nn\\ 
  b^{\mu}_{0}(t)=\sum^{\infty}_{m=1}b^{\mu}_{0m}\sqrt{\frac{2}{T}}\sin(\frac{\pi mt}{T})
  \eea
\bea
 \partial_{t}a_{n}(t)=\sum^{\infty}_{m=1}a^{\mu}_{nm}\sqrt{\frac{2}{T}}\cos(\frac{\pi mt}{T})\cdot\frac{\pi m}{T}\nn\\
 \partial_{t}b_{n}(t)=\sum^{\infty}_{m=1}b^{\mu}_{nm}\sqrt{\frac{2}{T}}\cos(\frac{\pi mt}{T})\cdot\frac{\pi m}{T}\nn\\
 \partial_{t}b_{0}(t)=\sum^{\infty}_{m=1}b^{\mu}_{0m}\sqrt{\frac{2}{T}}\cos(\frac{\pi mt}{T})\cdot\frac{\pi m}{T}
   \eea

For the integrals on $t$, one obtains: 
\bea
\int^{T}_{0}dt(a^{\mu}_{n}(t))^{2}=\sum^{\infty}_{m=1}(a^{\mu}_{nm})^{2}\nn\\
\int^{T}_{0}dt(b^{\mu}_{n}(t))^{2}=\sum^{\infty}_{m=1}(b^{\mu}_{nm})^{2}
\eea
\bea
\int^{T}_{0}dt(\partial_{t}a^{\mu}_{n}(t))^{2}=\sum^{\infty}_{m=1}(a^{\mu}_{nm})^{2}\frac{\pi^{2}m^{2}}{T^{2}} \nn\\
\int^{T}_{0}dt(\partial_{t}b^{\mu}_{n}(t))^{2}=\sum^{\infty}_{m=1}(b^{\mu}_{nm})^{2}\frac{\pi^{2}m^{2}}{T^{2}} \nn\\
\int^{T}_{0}dt(\partial_{t}b^{\mu}_{0}(t))^{2}=\sum^{\infty}_{m=1}(b^{\mu}_{0m})^{2}\frac{\pi^{2}m^{2}}{T^{2}} 
\eea
The part of $a^{\mu}$ in $A[X^{\mu}_{qu}]$, eq.(2.18):
\beq
\sum^{\infty}_{n=1}\sum^{\infty}_{m=1}[4\pi^{2}n^{2}+\frac{\pi^{2}m^{2}}{T^{2}}](a^{\mu}_{nm})^{2}
\eeq
The part of $b^{\mu}$ in $A[X^{\mu}_{qu}]$, eq.(2.18):
\beq
\sum^{\infty}_{n=1}\sum^{\infty}_{m=1}[4\pi^{2}n^{2}+\frac{\pi^{2}m^{2}}{T^{2}}](b^{\mu}_{nm})^{2}
\eeq
The part of $b_{0}^{\mu}$ in $A[X^{\mu}_{qu}]$, eq.(2.18):
\beq
\sum^{\infty}_{m=1}\frac{\pi^{2}m^{2}}{T^{2}}(b_{0m}^{\mu})^{2}
\eeq
Altogether, by the eq.(2.18), and the sum of the equations (2.23), (2.24), (2.25) we obtain:
\bea
A[X_{qu}]=\sum^{\infty}_{n=1}\sum^{\infty}_{m=1}[4\pi^{2}n^{2}+\frac{\pi^{2}m^{2}}{T^{2}}](a^{\mu}_{nm})^{2}\nn\\
+(4\pi^{2}n^{2}+\frac{\pi^{2}m^{2}}{T^{2}})^{2}(b^{\mu}_{nm})^{2}]+\sum^{\infty}_{m=1}\frac{\pi^{2}m^{2}}{T^{2}}(b^{\mu}_{0m})^{2}
\eea
The propagator of $X^{\mu}_{qu}$, for one $\mu$ component:
\bea
G_{X}(R;T)=\int DX_{qu} \exp(-A[X_{qu}])\nn\\
=\prod^{\infty}_{n=1}\prod^{\infty}_{m=1}(\int^{+\infty}_{-\infty}d a_{nm}\int^{+\infty}_{-\infty}db_{nm})\times\prod^{\infty}_{m=1}\int^{+\infty}_{-\infty}db_{0m}\nn\\
\times\prod^{\infty}_{n=1}\prod^{\infty}_{m=1}\exp\{-(4\pi^{2}n^{2}+\frac{\pi^{2}m^{2}}{T^{2}})a^{2}_{nm}\}\nn\\
\times\prod^{\infty}_{n=1}\prod^{\infty}_{m=1}\exp\{-(4\pi^{2}n^{2}+\frac{\pi^{2}m^{2}}{T^{2}})b^{2}_{nm}\}\nn\\
\times\prod^{\infty}_{m=1}\exp\{-\frac{\pi^{2}m^{2}}{T^{2}}b^{2}_{0m}\}\nn\\
\propto\prod^{\infty}_{n=1}\prod^{\infty}_{m=1}\frac{1}{4\pi^{2}n^{2}+\frac{\pi^{2}m^{2}}{T^{2}}}\times\prod^{\infty}_{m=1}\frac{T}{\pi m}
\eea
\beq
G_{X}(R;T)
\propto\prod^{\infty}_{n=1}\prod^{\infty}_{m=1}\frac{1}{4\pi^{2}n^{2}+\frac{\pi^{2}m^{2}}{T^{2}}}\times\prod^{\infty}_{m=1}\frac{T}{\pi m} 
\propto\prod^{\infty}_{n=1}\prod^{\infty}_{m=1}\frac{1}{4n^{2}+\frac{m^{2}}{T^{2}}}\times\prod^{\infty}_{m=1}\frac{T}{m} 
\eeq
One obtains from eq.(2.28):
\beq
G_{X}(R;T)\propto \frac{1}{\sqrt{T}}\frac{\exp\{\frac{\pi}{6}T\}}{\prod^{\infty}_{n=1}(1-e^{-4\pi n T})} 
\eeq
The formulas which have been used:
\beq
\prod^{\infty}_{m=1}\frac{T}{m}\propto\frac{1}{\sqrt{T}}
\eeq
\beq
\prod^{\infty}_{n=1}\prod^{\infty}_{m=1}\frac{1}{4n^{2}+\frac{m^{2}}{T^{2}}}
\propto e^{\frac{\pi T}{6}}\prod^{\infty}_{n=1}\frac{1}{1-e^{-4\pi nT}}
\eeq
In calculation of the products in (2.30), (2.31) the Riemann $\zeta(z)$ function have been used. The details are given in the Appendix A.

\vskip1cm

{\bf 2.} 

The ghosts part of the correlation function (2.3), with the propagator:
\bea
C_{b,c}(R;T)=\int Db(t,x)\int D\bar{b}(t,x)\int D c(t;x)\int D\bar{c}(t;x)\nn\\
\exp\{-B[b,\bar{b},c,\bar{c}]\}
\eea
The action $B(b,\bar{b},c,\bar{c})$, eq.(2.5), can be expressed in real components:
\bea
B(b,\bar{b},c,\bar{c})=\int^{T}_{0}dt\int^{1}_{0}dx(b\bar{\partial}c+\bar{b}\partial
\bar{c})\nn\\
=\int^{T}_{0}dt\int^{1}_{0}dx[b_{1}(\partial_{1}c_{1}-\partial_{2}c_{2})-b_{2}(\partial_{2}c_{1}+\partial_{1}c_{2})]
\eea
We have used the expressions:
\bea
b=b_{1}+ib_{2},\,\,\,\bar{b}=b_{1}-ib_{2},\,\,\,c=c_{1}+ic_{2},\,\,\,\bar{c}=c_{1}-ic_{2}\nn\\
\bar{\partial}=\frac{1}{2}(\partial_{1}+i\partial_{2}),\,\,\,\partial=\frac{1}{2}(\partial_{1}-i\partial_{2})
\eea
The decompositions of the fields $b_{1}, b_{2}, c_{1}, c_{2}$, being periodic in $x$ with the period $0<x<1$, will be taken of the form:
\bea
b_{1}(x,t)=\sum^{\infty}_{n=1}[b_{1,n}(t)\sqrt{2}\cos(2\pi nx)+\tilde{b}_{1,n}(t)\sqrt{2}\sin(2\pi nx)+b_{1,0}(t)\nn\\
c_{1}(x,t)=\sum^{\infty}_{n=1}[c_{1,n}(t)\sqrt{2}\sin(2\pi nx)+\tilde{c}_{1,n}(t)\sqrt{2}\cos(2\pi nx)+\tilde{c}_{1,0}(t)\nn\\
b_{2}(x,t)=\sum^{\infty}_{n=1}[b_{2,n}(t)\sqrt{2}\cos(2\pi nx)+\tilde{b}_{2,n}(t)\sqrt{2}\sin(2\pi nx)+b_{2,0}(t)\nn\\
c_{2}(x,t)=\sum^{\infty}_{n=1}[c_{2,n}(t)\sqrt{2}\sin(2\pi nx)+\tilde{c}_{2,n}(t)\sqrt{2}\cos(2\pi nx)+\tilde{c}_{2,0}(t)
\eea
The derivatives of the fields $c_{1}(x,t),\,\,c_{2}(x,t)$, 
which appear in the eq.(2.33), will be given by:
\bea
\partial_{1}c_{1}(x,t)=\partial_{x}c_{1}(x,t)\nn\\
=\sum^{\infty}_{n=1}[c_{2,n}(t)\sqrt{2}\cos(2\pi nx)-\tilde{c}_{1,n}(t)\sqrt{2}\sin(2\pi nx)]2\pi n\nn\\
\partial_{1}c_{2}(x,t)=\partial_{x}c_{2}(x,t)\nn\\
=\sum^{\infty}_{n=1}[c_{2,n}(t)\sqrt{2}\cos(2\pi nx)-\tilde{c}_{2,n}(t)\sqrt{2}\sin(2\pi nx)]2\pi n\nn\\
\partial_{2}c_{1}(x,t)=\partial_{t}c_{1}(x,t)\nn\\
=\sum^{\infty}_{n=1}[\partial_{t}c_{1,n}(t)\sqrt{2}\sin(2\pi nx)+\partial_{t}\tilde{c}_{1,n}(t)\sqrt{2}\cos(2\pi nx)]+\partial_{t}\tilde{c}_{1,0}(t)\nn\\
\partial_{2}c_{2}(x,t)=\partial_{t}c_{2}(x,t)\nn\\
=\sum^{\infty}_{n=1}[\partial_{t}c_{2,n}(t)\sqrt{2}\sin(2\pi nx)+\partial_{t}\tilde{c}_{2,n}(t)\sqrt{2}\cos(2\pi nx)]+\partial_{t}\tilde{c}_{2,0}(t)
\eea
In the Action $B(b,\bar{b},c,\bar{c})$, eq.(2.33), when integrated on $x$, appear four terms:
\bea
\int^{1}_{0}dx\,\,b_{1}\partial_{1}c_{1}=\sum^{\infty}_{n=1}[b_{1,n}(t)c_{1,n}(t)-\tilde{b}_{1,n}(t)\tilde{c}_{1,n}(t)]2\pi n\nn\\
-\int^{1}_{0}dx\,\,b_{1}\partial_{2}c_{2}=-\sum^{\infty}_{n=1}[b_{1,n}\partial_{t}\tilde{c}_{2,n}(t)+\tilde{b}_{1,n}\partial_{t}c_{2,n}(t)]-b_{1,0}(t)\partial_{t}\tilde{c}_{2,0}(t)\nn\\
-\int^{1}_{0}dx\,\,b_{2}\partial_{2}c_{1}=-\sum^{\infty}_{n=1}[b_{2,n}\partial_{t}\tilde{c}_{1,n}(t)+\tilde{b}_{2,n}(t)\partial_{t}c_{1,n}(t)]-b_{2,0}(t)\partial_{t}\tilde{c}_{1,0}(t)\nn\\
-\int^{1}_{0}dx\,\,b_{2}\partial_{1}c_{2}=\sum^{\infty}_{n=1}[-b_{2,n}(t)c_{2,n}(t)+\tilde{b}_{2,n}(t)\tilde{c}_{2,n}(t)]2\pi n
\eea

Next we decompose the fields in the above expressions in $t$ modes and we integrate them on $t$.

We start with the fields of the $x$--zero modes, $\tilde{c}_{1,0}(t)$, $\tilde{c}_{2,0}(t)$, $b_{1,0}(t)$, $b_{2,0}(t)$, with the terms, in (2.37),
\beq
-b_{1,0}(x)\partial_{t}\tilde{c}_{2,0}(t),\quad-b_{2,0}(t)\partial_{t}\tilde{c}_{1,0}(t)
\eeq
We take the following reduced decompositions on $t$ modes:
\bea
\tilde{c}_{1,0}(t)=\sum^{\infty}_{m=1}\tilde{c}_{1,0m}\sqrt{\frac{2}{T}}\sin(\frac{\pi mt}{T})\nn\\
\tilde{c}_{2,0}(t)=\sum^{\infty}_{m=1}\tilde{c}_{2,0m}\sqrt{\frac{2}{T}}\sin(\frac{\pi mt}{T})\nn\\
b_{1,0}(t)=\sum^{\infty}_{m=1}b_{1,0m}\sqrt{\frac{2}{T}}\cos(\frac{\pi mt}{T})\nn\\
b_{2,0}(t)=\sum^{\infty}_{m=1}b_{2,0m}\sqrt{\frac{2}{T}}\cos(\frac{\pi mt}{T})
\eea
\bea
\partial_{t}\tilde{c}_{1,0}(t)=\sum^{\infty}_{m=1}\tilde{c}_{1,0m}\sqrt{\frac{2}{T}}\cos(\frac{\pi mt}{T})\times\frac{\pi m}{T}\nn\\
\partial_{t}\tilde{c}_{2,0}(t)=\sum^{\infty}_{m=1}\tilde{c}_{2,0m}\sqrt{\frac{2}{T}}\cos(\frac{\pi mt}{T})\times\frac{\pi m}{T}
\eea
We put (2.39), (2.40) into (2.38) and we integrate them on $t$. We obtain:
\bea
-\int_{0}^{T}dt  b_{1,0}(x)\partial_{t}\tilde{c}_{2,0}(t)=\sum^{\infty}_{m=1}b_{1,0m}c_{2,0m}\times\frac{\pi m}{T}\nn\\
-\int_{0}^{T}dt  b_{2,0}(x)\partial_{t}\tilde{c}_{1,0}(t)=\sum^{\infty}_{m=1}b_{2,0m}\tilde{c}_{2,0m}\times\frac{\pi m}{T}
\eea

The $x$ zero modes part of the functional integral for the propagator is given by:
\bea
\int Db_{1,0}(t)\int D\tilde{c}_{2,0}(t)\int Db_{2,0}(t)\int D\tilde{c}_{1,0}(t)\nn\\
\times\prod^{\infty}_{m=1}\exp\{b_{1,0m}\tilde{c}_{1,0m}\frac{\pi m}{T}\}
\prod^{\infty}_{m=1}\exp\{b_{2,0m}\tilde{c}_{1,om}\frac{\pi m}{T}\}\nn\\
=\prod^{\infty}_{m=1}\int db_{1.0m}\prod_{m=1}^{\infty}\int d\tilde{c}_{2.0m}\prod^{\infty}_{m=1}\int db_{2.0m}\prod^{\infty}_{m=1}\int d\tilde{c}_{1,m}\nn\\
\times\prod^{\infty}_{m=1}\exp\{b_{1,0m}\tilde{c}_{2,om}\frac{\pi m}{T}\}
\prod^{\infty}_{m=1}\exp\{b_{2,0m}\tilde{c}_{1,om}\frac{\pi m}{T}\}
\eea
One obtains, for the zero modes part of the propagator,
\beq
\prod^{\infty}_{m=1}(\frac{\pi m}{T})^{2}
\eeq

Next, for non-zero $x$ modes, remains the following expressions in eq.(2.37):
\bea
\sum^{\infty}_{n=1}[b_{1,n}(t)c_{1,n}(t)+\tilde{b}_{1,n}(t)\tilde{c}_{1.n}(t)]2\pi n\nn\\
-\sum^{\infty}_{n=1}[b_{1,n}(t)\partial_{t}\tilde{c}_{2,n}(t)+\tilde{b}_{1,n}(t)\partial_{t}c_{2.n}(t)]\nn\\
-\sum^{\infty}_{n=1}[b_{2,n}(t)\partial_{t}\tilde{c}_{1,n}(t)+\tilde{b}_{2,n}(t)\partial_{t}c_{1.n}(t)]\nn\\
-\sum^{\infty}_{n=1}[b_{2,n}(t)c_{2,n}(t)+\tilde{b}_{2,n}(t)\tilde{c}_{2.n}(t)]2\pi n
\eea

For non-zero $x$ modes we take the following reduced expansions on $t$ modes, 
similar to those for the zero $x$ modes in eq.(2.39):
\bea
c_{1,n}(t)=\sum^{\infty}_{m=1}c_{1,nm}\sqrt{\frac{2}{T}}\cos(\frac{\pi mt}{T})\nn\\
c_{2,n}(t)=\sum^{\infty}_{m=1}c_{2,nm}\sqrt{\frac{2}{T}}\cos(\frac{\pi mt}{T})\nn\\
\tilde{c}_{1,n}(t)=\sum^{\infty}_{m=1}\tilde{c}_{1,nm}\sqrt{\frac{2}{T}}\sin(\frac{\pi mt}{T})\nn\\
\tilde{c}_{2,n}(t)=\sum^{\infty}_{m=1}\tilde{c}_{2,nm}\sqrt{\frac{2}{T}}\sin(\frac{\pi mt}{T})\nn\\
b_{1,n}(t)=\sum^{\infty}_{m=1}b_{1,nm}\sqrt{\frac{2}{T}}\cos(\frac{\pi mt}{T})\nn\\
b_{2,n}(t)=\sum^{\infty}_{m=1}b_{2,nm}\sqrt{\frac{2}{T}}\cos(\frac{\pi mt}{T})\nn\\
\tilde{b}_{1,n}(t)=\sum^{\infty}_{m=1}\tilde{b}_{1,nm}\sqrt{\frac{2}{T}}\sin(\frac{\pi mt}{T})\nn\\
\tilde{b}_{2,n}(t)=\sum^{\infty}_{m=1}\tilde{b}_{2,nm}\sqrt{\frac{2}{T}}\sin(\frac{\pi mt}{T})
\eea
We integrate on $t$ the expression (2.44), with the decomposition on $t$ modes 
in (2.45). We get:
\bea
\sum^{\infty}_{n=1}\sum^{\infty}_{m=1}[(b_{1,nm}c_{1,nm}+\tilde{b}_{1,nm}\tilde{c}_{1,nm})2\pi n\nn\\
-(b_{1,nm}\tilde{c}_{2,nm}+\tilde{b}_{1,nm}c_{2,nm})\frac{\pi m}{T}\nn\\
-(b_{2,nm}\tilde{c}_{1,nm}+\tilde{b}_{2,nm}c_{1,nm})\frac{\pi m}{T}\nn\\
-(b_{2,nm}c_{2,nm}+\tilde{b}_{2,nm}\tilde{c}_{2,nm})2\pi n]
\eea
Above is the action for ghosts, for non-zero $x$ modes.

The propagator, the non-zero $x$ modes part of ghosts, is obtained 
by the functional integral of the exponential of (2.46):
\bea
\prod^{\infty}_{n=1}\prod^{\infty}_{m=1}[\int db_{1,nm}\int d\tilde{b}_{1,nm}\int db_{2,nm}\int d\tilde{b}_{2,nm}\nn\\
\int d c_{1,nm}\int d\tilde{c}_{1,nm}\int dc_{2,nm}\int d\tilde{c}_{2,nm}]\nn\\
\prod^{\infty}_{n=1}\prod^{\infty}_{m=1}
\exp\{-(b_{1,nm}c_{1,nm}+\tilde{b}_{1,nm}\tilde{c}_{1,nm})2\pi n\nn\\
+(b_{1,nm}\tilde{c}_{2,nm}+\tilde{b}_{1,nm}c_{2,nm})\frac{\pi m}{T}\nn\\
+(b_{2,nm}\tilde{c}_{1,nm}+\tilde{b}_{2,nm}c_{1,nm})\frac{\pi m}{T}\nn\\
+(b_{2,nm}c_{2,nm}+\tilde{b}_{2,nm}\tilde{c}_{2,nm})2\pi n\}
\eea
One obtains, after integration:
\bea
\prod^{\infty}_{n=1}\prod^{\infty}_{m=1}(16\pi^{4}n^{4}+8\pi^{4}n^{2}\frac{m^2}{T^{2}}+\frac{\pi^{4}m^{4}}{T^{4}})\nn\\
=\prod^{\infty}_{n=1}\prod^{\infty}_{m=1}(4\pi^{2}n^{2}+\frac{\pi^{2}m^{2}}{T^{2}})^{2}
\eea
For the zero $x$ modes we have obtained the expression in the eq.(2.43).
Together, we find the following expression for the ghosts propagator:
\beq
G_{b,c}(R;T)=(\prod^{\infty}_{m=1}\frac{\pi m}{T})^{2}\times
(\prod^{\infty}_{n=1}\prod^{\infty}_{m=1}(4\pi^{2}n^{2}+\frac{\pi^{2}m^{2}}{T^{2}}))^{2}
\eeq

Earlier we have obtained, eq.(2.28), when the product over the $\mu$ components is taken, the following expression for the propagator of the $X^{\mu}_{qu}(t,x)$ field:
\beq
G_{X}(R;T)=\prod^{26}_{\mu=1}\int D X^{\mu}_{qu}e^{-A[X^{\mu}_{qu}]}=(\prod^{\infty}_{m=1}\frac{T}{\pi m})^{26}(\prod^{\infty}_{n=1}\prod^{\infty}_{m=1}\frac{1}{(4\pi^{2}n^{2}+\frac{\pi^{2}m^{2}}{T^{2}})})^{26}
\eeq 
We can conclude that the ghosts compensate, exactly, the contribution to the propagator of two components of the $X^{\mu}_{qu}$ fields, the two non-physical fluctuations of this field. 
In particular, this justifies the reduced decompositions, on $t$ - modes, (2.39), (2.45) of the ghosts. The correlation function $<p(R)p(0)>$ is defined by the equation (2.3). We obtain:
\bea
<p(R)p(0)>=\int^{\infty}_{0}dT e^{-\frac{1}{T}R^{2}}(\prod^{\infty}_{m=1}\frac{T}{\pi m})^{24}\times((\prod^{\infty}_{n=1}\prod^{\infty}_{m=1}\frac{1}{(4\pi^{2}n^{2}+\frac{\pi^{2}m^{2}}{T})^{2}})^{24}\nn\\
\propto\int^{\infty}_{0}dT e^{-\frac{R^{2}}{T}}(\frac{1}{\sqrt{T}})^{24}(\frac{\exp\{\frac{\pi T}{6}\}}{(1-e^{-4\pi nT})})^{24} \nn\\
=\int^{\infty}_{0}dT e^{-\frac{1}{T}R^{2}}\frac{1}{T^{12}}\times\frac{\exp\{4\pi T\}}{\prod^{\infty}_{n=1}(1-e^{-4\pi nT})^{24}}
\eea

We have reobtained the well known result [1].

The expression (2.51) is highly divergent at $T\rightarrow\infty$. It is pathological, in case of the bosonic string.

In the next Section we shall define the correlation function $<p(R)p(0)>$ for the fermionic string, in its critical dimension $D=10$. In that case we shall find the function which is well defined, with the $T$ integral convergent.

\section{Fermion string two-point function.}

Correlation function of the fermionic string punched at points $\vec{X}_1=\vec{0}$ 
and $\vec{X}_2=\vec{R}$ of the $D$ dimensional space $(D=10)$ 
is  given by the general expression:
\beq
<p(R)p(0)> = \int^{\infty}_{0}dT\,\,G(R;T)
\eeq
\bea
G(R;T)=\int DX^{\mu}(t,x)\int D\psi^{\mu}(t,x)\int D\bar{\psi}^{\mu}(t,x)\nn\\
\int D b(t,x)\int D\bar{b}(t,x)\int D c(t,x)\int D\bar{c}(t,x)\nn\\
\int D\beta(t,x) \int D\gamma(t,x)\nn\\
\times \exp\{-A_{X}[X^{\mu}]-A_{\psi,\psi}[\psi^{\mu},\bar{\psi}^{\mu}]\nn\\
-A_{b,c}[b,\bar{b},c,\bar{c}]-A_{\beta,\gamma}[\beta,\gamma]\}
\eea
We have seen in Sec.2 that, for our function, the role of the ghosts, $b,\bar{b},c,\bar{c}$ 
is just to compensate for two component of $X^{\mu}$.

It will be similar for our present function in (3.1), (3.2). We shall suppose that the role of the ghosts $\beta,\gamma$ in (3.1), (3.2) will be similar, 
just to compensate for two components of the fermion fields $\psi^{\mu}$ and $\bar{\psi}^{\mu}$. Then our expression for the propagator (3.2) can be taken in the reduced form:
\bea
G(R;T)=\int DX^{\mu}(t,x)\int D\bar{\psi}^{\mu}(t,x)\nn\\
\times\exp\{-A_{X}[X^{\mu}]-A_{\psi,\bar{\psi}}[\psi^{\mu},\bar{\psi}^{\mu}]\}
\eea
with $\mu=1,...,8$ both for $X^{\mu}$ and for $\psi^{\mu}$, $\bar{\psi}^{\mu}$ 
fields in (3.3), 
with the actions:
\beq
A_{X}[X^{\mu}]=\int^{T}_{0}dt[(\partial_{t}X^{\mu})^{2}+(\partial_{x}X^{\mu})^{2}]
\eeq
\beq
A_{\psi,\bar{\psi}}[\psi^{\mu},\bar{\psi}^{\mu}]=\int^{T}_{0}dt\int^{1}_{0}dx(\psi^{\mu}\bar{\partial}\psi^{\mu}+\bar{\psi}^{\mu}\partial\bar{\psi}^{\mu})
\eeq
The $\mu$ components of $X^{\mu}$ and $\psi^{\mu}$, $\bar{\psi}^{\mu}$ 
are independent for our correlation functions. For one $\mu$ component: 
\beq
A_{X}[X]=\int^{T}_{0}dt\int^{1}_{0}dx[(\partial_{t}X)^{2}+(\partial_{x}X)^{2}]
\eeq
\beq
A_{\psi.\bar{\psi}}[\psi,\bar{\psi}]=\int^{T}_{0}dt\int^{1}_{0}dx(\psi\bar{\partial}\psi+\bar{\psi}\partial\bar{\psi})
\eeq

For the $X$ part of the correlation function the calculations have already been done in the Sec.2, with the results:
\bea
G_{X}(R;T)=e^{-\frac{R^{2}}{T}}\times\prod^{\infty}_{m=1}\frac{T}{\pi m}\times\prod^{\infty}_{n=1}\prod^{\infty}_{m=1}\frac{1}{(4\pi^{2}n^{2}+\frac{\pi^{2}m^{2}}{T^{2}})}\nn\\
\propto e^{-\frac{1}{T}R^{2}}\times\frac{1}{\sqrt{T}}\frac{\exp\{\frac{\pi T}{6}\}}{\prod^{\infty}_{n=1}(1-e^{-4\pi nT})}
\eea
-- equations (2.28), (2.29), Sec.2.

We shall now calculate the $\psi$, $\bar{\psi}$ part of the propagator $G(R;T)$, 
equations (3.3) -- (3.5):
\bea
G_{\psi,\bar{\psi}}=\int D\psi(t,x)D\bar{\psi}(t,x)\exp\{-A_{\psi,\bar{\psi}}[\psi,\bar{\psi}]\}\nn\\
=\int D\psi(t,x)\int D\bar{\psi}(t,x)\exp\{-\int^{T}_{0}dt\int^{1}_{0}dx(\psi\bar{\partial}\psi+\bar{\psi}\partial\bar{\psi})\}
\eea
We shall give the real form for the action $A_{\psi,\bar{\psi}}[\psi,\bar{\psi}]$, eq.(3.7).
\bea
\psi=\psi_{1}+i\psi_{2},\quad\bar{\psi}=\psi_{1}-i\psi_{2}\nn\\
\bar{\partial}=\frac{1}{2}(\partial_{1}+i\partial_{2}),\quad\partial=\frac{1}{2}(\partial_{1}-i\partial_{2}),\nn\\
\partial_{1}=\partial_{x},\quad\partial_{2}=\partial_{t}
\eea
$\psi_{1}$, $\psi_{2}$ above are real. We find
\beq
A_{\psi,\bar{\psi}}[\psi,\bar{\psi}]=\int^{T}_{0}dt\int^{1}_{0}dx(\psi_{1}\partial_{1}\psi_{1}-\psi_{1}\partial_{2}\psi_{2}-\psi_{2}\partial_{2}\psi_{1}-\psi_{2}\partial_{1}\psi_{2})
\eeq
We shall denote as $I_{\psi,\bar{\psi}}$ the $\psi,\bar{\psi}$ part of the propagator:
\bea
I_{\psi,\bar{\psi}}=\int D\psi_{1}\int D\psi_{2}\exp\{-A_{\psi,\bar{\psi}}[\psi,\bar{\psi}]\}\nn\\
=\int D\psi_{1}\int D\psi_{2}\exp\{-\int^{T}_{0}dt\int^{1}_{0}dx(\psi_{1}\partial_{1}\psi_{1}-\psi_{1}\partial_{2}\psi_{2}-\psi_{2}\partial_{2}\psi_{1}-\psi_{2}\partial_{1}\psi_{2})\}
\eea
The components $\psi_{1}(t,x)$, $\psi_{2}(t,x)$ are periodic in $x$, $x\rightarrow x+1$. They can be decomposed as follows:
\bea
\psi_{1}(t,x)=\sum^{\infty}_{n=1}[\psi_{1,n}(t)\sqrt{2}\sin(2\pi nx)+\tilde{\psi}_{1,n}(t)\sqrt{2}\cos(2\pi nx)]+\tilde{\psi}_{1,0}(t)\nn\\
\psi_{2}(t,x)=\sum^{\infty}_{n=1}[\psi_{2,n}(t)\sqrt{2}\sin(2\pi nx)+\tilde{\psi}_{2,n}(t)\sqrt{2}\cos(2\pi nx)]+\tilde{\psi}_{2,0}(t)
\eea
Their derivatives:
\bea
\partial_{1}\psi_{1}(t,x)=\sum^{\infty}_{n=1}[\psi_{1,n}(t)\sqrt{2}\cos(2\pi nx)-\tilde{\psi}_{1,n}(t)\sqrt{2}\sin(2\pi nx)]2\pi n\nn\\
\partial_{1}\psi_{2}(t,x)=\sum^{\infty}_{n=1}[\psi_{2,n}(t)\sqrt{2}\cos(2\pi nx)-\tilde{\psi}_{2,n}(t)\sqrt{2}\sin(2\pi nx)]2\pi n\nn\\
\partial_{2}\psi_{1}(t,x)=\sum^{\infty}_{n=1}[\partial_{t}\psi_{1,n}(t)\sqrt{2}\sin(2\pi nx)+\partial_{t}\tilde{\psi}_{1,n}(t)\sqrt{2}\cos(2\pi nx)]+\partial_{t}\tilde{\psi}_{1,0}(t)\nn\\
\partial_{2}\psi_{2}(t,x)=\sum^{\infty}_{n=1}[\partial_{t}\psi_{2,n}(t)\sqrt{2}\sin(2\pi nx)+\partial_{t}\tilde{\psi}_{2,n}(t)\sqrt{2}\cos(2\pi nx)]+\partial_{t}\tilde{\psi}_{2,0}(t)
\eea
We calculate next $A_{\psi,\bar{\psi}}[\psi,\bar{\psi}]$, eq.(3.11), by integrating
for the moment on $x$ alone. We find:
\bea
A_{\psi,\bar{\psi}}[\psi,\bar{\psi}]=\int^{T}_{0}dt\{\sum^{\infty}_{n=1}[-2\psi_{1,n}(t)\tilde{\psi}_{1,n}(t)\times2\pi n\nn\\
-\psi_{1,n}(t)\partial_{t}\psi_{2,n}(t)-\tilde{\psi}_{1,n}(t)\partial_{t}\psi_{2,n}(t)\nn\\
-\psi_{2,n}(t)\partial_{t}\psi_{1,n}(t)-\tilde{\psi}_{2,n}(t)\partial_{t}\psi_{1,n}(t)\nn\\
+2\psi_{2,n}(t)\tilde{\psi}_{2,n}(t)\times 2\pi n]\nn\\
-\tilde{\psi}_{1,0}(t)\partial_{t}\tilde{\psi}_{2,0}(t)-\tilde{\psi}_{2,0}(t)\partial_{t}\tilde{\psi}_{1,0}(t)\}
\eea
We decompose now $\psi_{1,n}(t)$, $\psi_{2,n}(t)$, $\tilde{\psi}_{1,n}(t)$, 
$\tilde{\psi}_{2,n}(t)$ on $t$ modes. First we do it for the non-zero $x$ modes. We take the following reduced decompositions:
\bea 
\psi_{1,n}(t)=\sum^{\infty}_{m=1}\psi_{1,nm}\sqrt{\frac{2}{T}}\sin(\frac{\pi mt}{T})\nn\\
\psi_{2,n}(t)=\sum^{\infty}_{m=1}\psi_{2,nm}\sqrt{\frac{2}{T}}\cos(\frac{\pi mt}{T})\nn\\
\tilde{\psi}_{1,n}(t)=\sum^{\infty}_{m=1}\tilde{\psi}_{1,nm}\sqrt{\frac{2}{T}}\sin(\frac{\pi mt}{T})\nn\\
\tilde{\psi}_{2,n}(t)=\sum^{\infty}_{m=1}\tilde{\psi}_{2,nm}\sqrt{\frac{2}{T}}\cos(\frac{\pi mt}{T})
\eea
Their derivatives:
\bea
\partial_{t}\psi_{1,n}(t)=\sum^{\infty}_{m=1}\psi_{1,nm}\sqrt{\frac{2}{T}}\cos(\frac{\pi mt}{T})\times(\frac{\pi m}{T})\nn\\
\partial_{t}\psi_{2,n}(t)=\sum^{\infty}_{m=1}\psi_{2,nm}\sqrt{\frac{2}{T}}\sin(\frac{\pi mt}{T})\times(-\frac{\pi m}{T})\nn\\
\partial_{t}\tilde{\psi}_{1,n}(t)=\sum^{\infty}_{m=1}\tilde{\psi}_{1,nm}\sqrt{\frac{2}{T}}\cos(\frac{\pi mt}{T})\times(\frac{\pi m}{T})\nn\\
\partial_{t}\tilde{\psi}_{2,n}(t)=\sum^{\infty}_{m=1}\tilde{\psi}_{2,nm}\sqrt{\frac{2}{T}}\sin(\frac{\pi mt}{T})\times(-\frac{\pi m}{T})
\eea
Next we integrate on $t$ the action $A_{\psi\bar{\psi}}(\psi,\bar{\psi})$ in eq.(3.15),
its non zero $x$ modes part.

We find:
\bea
(A_{\psi\bar{\psi}}(\psi,\bar{\psi}))_{n.z.} = \sum^{\infty}_{n=1}\sum^{\infty}_{m=1}\{-2\psi_{1,nm}\tilde{\psi}_{1,nm}\times 2\pi n\nn\\
+(\psi_{1,nm}\psi_{2,nm}+\tilde{\psi}_{1,nm}\tilde{\psi}_{2,nm})\times\frac{\pi m}{T}\nn\\
-(\psi_{2,nm}\psi_{1,nm}+\tilde{\psi}_{2,nm}\tilde{\psi}_{1,nm})\times\frac{\pi m}{T}\nn\\
+2\psi_{2,nm}\tilde{\psi}_{2,nm}\times 2\pi n\}\nn\\
=\sum^{\infty}_{n=1}\sum^{\infty}_{m=1}\{(-2\psi_{1,nm}\tilde{\psi}_{1,nm}+2\tilde{\psi}_{2,nm}\tilde{\psi}_{2,nm})2\pi n\nn\\
+(2\psi_{1,nm}\psi_{2,nm}+2\tilde{\psi}_{1,nm}\tilde{\psi}_{2,nm})\times\frac{\pi m}{T}\}
\eea

Next,
\bea
 (I_{\psi,\bar{\psi}})_{n.z.}=\int D\psi\int D\bar{\psi}\exp\{-A[\psi,\bar{\psi}]\}\nn\\
=\prod^{\infty}_{n=1}\prod^{\infty}_{m=1}
\int d\psi_{1,nm} \int d\psi_{2,nm}\int d\tilde{\psi}_{1,nm}\int d\tilde{\psi}_{2,nm}\nn\\
\times\prod^{\infty}_{n=1}\prod^{\infty}_{m=1}
\exp\{(2\psi_{1,nm}\tilde{\psi}_{1,nm}-2\psi_{2,nm}\tilde{\psi}_{2,nm})2\pi n\nn\\
-(2\psi_{1,nm}{\psi}_{2,nm}+2\tilde{\psi}_{1,nm}\tilde{\psi}_{2,nm})\frac{\pi m}{T}\}\nn\\
=\prod^{\infty}_{n=1}\prod^{\infty}_{m=1}(16\pi^{2}n^{2}+4\frac{\pi m^{2}}{T^{2}})
=\prod^{\infty}_{n=1}\prod^{\infty}_{m=1}4(4\pi^{2}n^{2}+\frac{\pi^{2}m^{2}}{T^{2}}),
\eea
We have obtained $I_{\psi,\bar{\psi}}$, non-zero $x$ modes,
\beq
(I_{\psi,\bar{\psi}})_{n.z.}=\prod^{\infty}_{n=1}\prod^{\infty}_{m=1}4(4\pi^{2}n^{2}+\frac{\pi^{2}m^{2}}{T^{2}})
\eeq
We remind that:
$I_{X}$, non-zero $x$ modes,
\beq
(I_{X})_{n.z.}=\prod^{\infty}_{n=1}\prod^{\infty}_{m=1} 
\frac{1}{(4\pi^{2}n^{2}+\frac{\pi^{2}m^{2}}{T^{2}})}
\eeq
-- eq.(3.8).

We find that
\beq
(I_{X})_{n.z.} \times (I_{\psi,\bar{\psi}})_{n.z.} \propto 1
\eeq

We have to treate now the zero $x$ modes of the fermions, $\psi$, $\bar{\psi}$.

We shall consider the correlation function $<p(R)p(0)>$ of the fermionic  string in which the $x$ zero modes of the fermions $\tilde{\psi}_{1,0}(t)$, $\tilde{\psi}_{2,0}(t)$, in eq.(3.15), have fixed values at $t=0$ and at $t=T$, which means at $\vec{0}$, and at $\vec{R}$ 
in the $\vec{X}$ space. This corresponds, in particular, to the limiting case of two small loops, at points $\vec{0}$ and $\vec{R}$, the loops with the fermions having fixed values at these loops, the loops which tend to points, at $\vec{0}$ and at $\vec{R}$.

The correlation function of this type corresponds to the energy-energy correlation function $<\varepsilon(\vec{R})\varepsilon(\vec{0})>$ of the 3D Ising model, in its presentation by the fermionic string, on the lattice [2] -- [6].

In our present calculations, at the critical dimension $D=10$, we are very 
far from the non-critial $3D$ fermionic string. But still we consider, at $D=10$, the object which is similar to the $<\varepsilon(R)\varepsilon(0)>$ correlation function. 
This is for a vague motivation.

The action for the zero $x$ modes of $\tilde{\psi}_{1}$, $\tilde{\psi}_{2}$ is given by eq.(3.15): 
\beq
(A[\psi,\bar{\psi}])_{0}=-\int_{0}^{T}dt[\tilde{\psi}_{1,0}(t)\partial_{t}\tilde{\psi}_{2.0}(t)+\tilde{\psi}_{2,0}(t)\partial_{t}\tilde{\psi}_{1,0}(t)]
\eeq
We take $\tilde{\psi}_{1,0}(t)$, $\tilde{\psi}_{2,0}(t)$ in the form:
\bea
\tilde{\psi}_{1,0}(t)=\Psi_{1,0}+\tilde{\tilde{\psi}}_{1,0}(t) \nn\\
\tilde{\psi}_{2,0}(t)=\Psi_{2,0}+\tilde{\tilde{\psi}}_{2,0}(t)
\eea
where $\Psi_{1,0}$, $\Psi_{2,0}$ are constantes 
and $\tilde{\tilde{\psi}}_{1,0}(t)$, $\tilde{\tilde{\psi}}_{2,0}(t)$ 
are having zero boundary values at $t=0$ and $t=T$:
\bea
\tilde{\tilde{\psi}}_{1,0}(0)=\tilde{\tilde{\psi}}_{1,0}(T)=0 \nn\\
\tilde{\tilde{\psi}}_{2,0}(0)=\tilde{\tilde{\psi}}_{2,0}(T)=0
\eea
$\tilde{\psi}_{1,0}(t)$ and $\tilde{\psi}_{2,0}(t)$ of this form, eq.(3.24), have fixed boundary values at $t=0$ and $t=T$:
\bea
\tilde{\psi}_{1,0}(0)=\tilde{\psi}_{1,0}(T)=\Psi_{1,0}\nn\\
\tilde{\psi}_{2,0}(0)=\tilde{\psi}_{2,0}(T)=\Psi_{2,0}
\eea
We put equations (3.24) into eq.(3.23) and we calculate the action (3.23) 
of $\tilde{\psi}_{1,0}(t)$, $\tilde{\psi}_{2,0}(t)$ in eq.(3.24).
\bea
(A(\psi,\bar{\psi}))_{0}=-\int^{T}_{0}dt
[(\Psi_{1,0}+\tilde{\tilde{\psi}}_{1,0}(t))\partial_{t}\tilde{\tilde{\psi}}_{2,0}\nn\\  
+(\Psi_{2,0}+\tilde{\tilde{\psi}}_{2,0}(t))\partial_{t}\tilde{\tilde{\psi}}_{1,0}(t)]\nn\\
=-\int_{0}^{T}dt[\tilde{\tilde{\psi}}_{1,0}(t)\partial_{2}\tilde{\tilde{\psi}}_{2,0}(t) 
+\tilde{\tilde{\psi}}_{2,0}(t)\partial_{t}\tilde{\tilde{\psi}}_{1,0}(t)]\nn\\
=-2\int_{0}^{T}dt\tilde{\tilde{\psi}}_{1,0}(t)\partial_{t}\tilde{\tilde{\psi}}_{2,0}(t)
\eea
Next, $\tilde{\tilde{\psi}}_{1,0}(t)$, $\tilde{\tilde{\psi}}_{2,0}(t)$ 
having zero boundary values 
at $t=0$ and  at $t=T$, eq.(3.25), they expand in sinus, $\{\sin(\frac{\pi mt}{T})$, $m=1,2,...\infty\}$. Then the derivatives $\partial_{t}\tilde{\tilde{\psi}}_{1,0}(t)$, $\partial_{t}\tilde{\tilde{\psi}}_{2,0}(t)$, they expand in cosinus,$\{\cos(\frac{\pi mt}{T})$, 
$m=1,2,...\infty\}$. In this case the integral in eq.(3.27) will always be zero. 
So that the action of zero $x$ modes, eq.(3.23), will always be zero 
for $\tilde{\psi}_{1,0}(t)$, $\tilde{\psi}_{2,0}(t)$ having fixed boundary conditions, eq.(3.26). 
In this case for the propagator $(I_{\psi},\bar{\psi})_{0}$ we obtain:
\beq
(I_{\psi,\bar{\psi}})_{0}=\exp\{-(A[\psi,\bar{\psi}])_0\}=1
\eeq

Now, on the whole, we have found
\beq
I_{X}\equiv G_{X}(R;T)=e^{-\frac{R^{2}}{T}}\prod_{m=1}^{\infty}\frac{T}{\pi m}\times\prod^{\infty}_{n=1}\prod_{m=1}^{\infty}\frac{1}{(4\pi^{2}n^{2}+\frac{\pi^{2}m^{2}}{T^{2}})}
\eeq
-- eq.(3.8),
\beq
I_{\psi,\bar{\psi}}=\prod^{\infty}_{n=1}\prod^{\infty}_{m=1}4(4\pi^{2}n^{2}+\frac{\pi^{2}m^{2}}{T^{2}})
\eeq
-- eq.(3.20).
Putting them together, we obtain
\beq
G(R;T)=I_{X}\times I_{\psi,\bar{\psi}}\propto e^{-\frac{R^{2}}{T}}\prod^{\infty}_{m=1}\frac{T}{\pi m}\propto e^{-\frac{R^{2}}{T}}\frac{1}{\sqrt{T}}
\eeq
This is the result for one $\mu$ component, $\mu=1,2,... D-2=1,2,...8$.
For $D-2$ $\mu$ components,
\beq
G(R;T)\propto e^{-\frac{R^{2}}{T}}\times\frac{1}{(T)^{\frac{D-2}{2}}}=e^{-\frac{R^{2}}{t}}\times\frac{1}{(T)^4}
\eeq 
For the correlation function $<p(R)p(0)>$, eq.(3.1),we obtain
\beq
<p(R)p(0)>=\int^{\infty}_{0}dT \,G(R;T)=\int^{\infty}_{0}dT \,e^{-\frac{R^{2}}{T}}\times\frac{1}{T^{4}} = \frac{2}{R^6}
\eeq
The result is simple but well defined, the $T$ integral is convergent.

\section{Conclusions.}

We have defined and calculated the correlation function of punctual states of the fermionic string theory, in its critical dimension $D=10$.

Compared to the bosonic string, the correlation function is no longer divergent, because of compensation between the bosonic and fermionic excitations. The correlation function is found to be simple and well defined.

Divergence of the $T$ integral for the bosonic string correlator signify that the surfaces 
of the propagation of the bosonic string (bosonic surfaces) degenerate into the long thin tubes. The bosonic surfaces are unstable in this respect.

This does not happen for the fermion string correlator. 
The surfaces of its propagation remain extended surfaces.

Back to the $3D$ Ising.

The self avoiding surfaces (the domain walls) of the partition sum 
for the $3D$ Ising model, they degenerate (their dominant configurations) 
into the branched polymers, 
the long thin tubes again, branched this time, instead of extended surfaces [9] -- [11].
The surfaces are self avoiding, but still they are bosonic, having the same unstability.

We would expect that the fermion string surfaces, of the $3D$ Ising model [2] -- [8], 
will be stable, will remain extended surfaces, 
like their partners in the critical dimension $D=10$.

\vskip1.5cm

{\bf Acknowledgments.}
 
I am grateful to Marco Picco for the discussions and supports, 
and to Valentina Dotsenko for much help in the preparation of the manuscript.

\appendix.

\section{Infinite products.}

{\bf 1)}

\beq
\prod^{\infty}_{m=1}\frac{T}{m}=\exp\{\sum^{\infty}_{m=1}\log\frac{T}{m}\}
\eeq
\beq
\sum^{\infty}_{m=1}\log\frac{T}{m}=\sum^{\infty}_{m=1}\log T+\sum^{\infty}_{m=1}\frac{1}{m}
\eeq
\beq
\sum^{\infty}_{m=1}\log T=\zeta(0)\log T=-\frac{1}{2}\log T=\log\frac{1}{\sqrt{T}}
\eeq
\bea
\sum^{\infty}_{m=1}\log\frac{1}{m}=\lim_{z\rightarrow 0}\sum^{\infty}_{m=1}\frac{1}{z}[(\frac{1}{m})^{z}-1]\nn\\
=\lim_{z\rightarrow 0}\frac{1}{z}[\sum^{\infty}_{m=1}(\frac{1}{m^{z}}-1)]\nn\\
=\lim_{z\rightarrow 0}\frac{1}{z}[\zeta(z)-\zeta(0)]=\zeta'(0)=-\frac{1}{2}\log 2\pi=\log\frac{1}{\sqrt{2\pi}}
\eea
Eq.(A.2):
\beq
\sum^{\infty}_{m=1}\log\frac{T}{m}=\log\frac{1}{\sqrt{T}}+\log\frac{1}{\sqrt{2\pi}}=\log\frac{1}{\sqrt{2\pi T}}
\eeq
Eq.(A.1):
\beq
\prod^{\infty}_{m=1}\frac{t}{m}=\exp\{\log\frac{1}{\sqrt{2\pi T}}\}=\frac{1}{\sqrt{2 \pi T}}
\eeq
\beq
\prod^{\infty}_{m=1}\frac{T^{2}}{m^{2}}=\frac{1}{2\pi T}
\eeq
Equation (A.6) could also be be factorized:
\beq
\prod^{\infty}_{m=1}\frac{1}{m}=\frac{1}{\sqrt{2\pi}}
\eeq
\beq
\prod^{\infty}_{m=1} T=\frac{1}{\sqrt{T}}
\eeq

\vskip1cm

{\bf 2)}

\bea
\prod^{\infty}_{n=1}\prod^{\infty}_{m=1}\frac{1}{4n^{2}+\frac{m^{2}}{T^{2}}}\nn\\
=\prod^{\infty}_{n=1}\prod^{\infty}_{m=1}(\frac{T^{2}}{m^{2}}\times\frac{1}{1+\frac{4n^{2}T^{2}}{m^{2}}})\nn\\
=\prod^{\infty}_{n=1}(\prod^{\infty}_{m-1}\frac{T^{2}}{m^{2}}\times\prod^{\infty}_{m=1}\frac{1}{1+\frac{4n^{2}T^{2}}{m^{2}}})
\eea
\beq
\prod^{\infty}_{m=1}\frac{T^{2}}{m^{2}}=\frac{1}{2 \pi T}
\eeq
\beq
\prod^{\infty}_{m=1}\frac{1}{1+\frac{4n^{2}T^{2}}{m^{2}}}=
(\prod^{\infty}_{m=1}(1+\frac{4n^{2}T^{2}}{m^{2}}))^{-1}
\eeq
we shall use the formula:
\beq
\prod^{\infty}_{m=1}(1+\frac{a^{2}}{m^{2}})=\frac{\sinh \pi a}{\pi a}=\frac{1}{2\pi a}(e^{\pi a}-e^{-\pi a}) = \frac{e^{\pi a}}{2\pi a}(1-e^{-2\pi a})
\eeq
We find:
\beq
\prod^{\infty}_{m=1}(1+\frac{4n^{2}T^{2}}{m^{2}})^{-1}
=[\frac{1}{2\pi\cdot 2 n T} e^{2\pi n T}(1-e^{-4\pi n T})]^{-1}
=4\pi n T\frac{e^{-2\pi n T}}{1-e^{-4\pi n T}}
\eeq
We put (A.11) and (A.14) into (A.10):
\bea
\prod^{\infty}_{n=1}\prod^{\infty}_{m=1}\frac{1}{4n^{2}+\frac{m^{2}}{T^{2}}}
=\prod^{\infty}_{n=1}(\frac{1}{2\pi T}\times 4\pi n T\frac{e^{-2\pi n T}}{1-e^{-4\pi n T}})\nn\\
=\prod^{\infty}_{n=1}2n\times \exp\{-2\pi T\sum^{\infty}_{n=1}n\} \times \frac{1}{\prod^{\infty}_{n=1}(1-e^{-4\pi n T})}
\eea
We shall use once again the $\zeta$ function value:
\beq
\sum^{\infty}_{n=1}n=\zeta(-1)=-\frac{1}{12}
\eeq
Finally:
\beq
\prod^{\infty}_{n=1}\prod^{\infty}_{m=1}\frac{1}{4n^{2}+\frac{m^{2}}{T^{2}}}
= \prod^{\infty}_{n=1}2n\times\frac{e^{\frac{\pi}{6}T}}{\prod^{\infty}_{n=1}(1-e^{-4\pi n T})}
\propto\frac{e^{\frac{\pi}{6}T}}{\prod^{\infty}_{n=1}(1-e^{-4\pi n T)}}
\eeq
We have used:
\beq
\prod^{\infty}_{n=1}2n=\prod^{\infty}_{n=1}2\times\prod^{\infty}_{n=1}n=\frac{1}{\sqrt{2}}\times\sqrt{2\pi}=\sqrt{\pi}\propto 1
\eeq


\begin{thebibliography}{99}


\bibitem{ref3} J.Polchinski, Preprint, unpublished.

\bibitem{ref1a}       A.M.Polyakov, 
                        Phys.Lett.{\bf 82B} (1979), 247.
             
                     
\bibitem{ref1b}       Vl.S.Dotsenko and A.M.Polyakov, 
                      Advanced Stadies in Pure Mathematics.  {\bf v.16} (1988), 171.                  

\bibitem{ref1c}       S.Samuel, 
                     J.Math. Phys.  {\bf 21} (1980) 2806,2815,2820.  



\bibitem{ref1e}     E.Fradkin, M.Srednicki and L.Susskind, 
                       Phys.Rev. {\bf D21} (1980) 2885.

\bibitem{ref1f}      A.Casher, D.Foerster and P. Windey,
                      Nucl.Phys. {\bf B251}, (1985) 29.


\bibitem{ref1}        C.Itzykson,
                     Nucl. Phys. {\bf B 210[B210]} (1982) 477.              

\bibitem{ref2}       Vl. S. Dotsenko 
                     Nucl. Phys.  {\bf B 285[FS19]} (1987) 45.

\bibitem{ref3}        Vl.S.Dotsenko, Paul Windey, Geoffrey Harris, Enzo Marinari,
                              Emil Martinec and Marco Picco,
                      Phys.Rev.Lett. 71,811(1993).
                              
\bibitem{ref4}          Vl.S.Dotsenko, Marco Picco, Paul Windey, Geoffrey Harris,
                                Emil Martinec, Enzo Marinari,
                       Nucl.Phys.{\bf B 488, p 577} (1995).
                                
\bibitem{ref5}            Vl.S.Dotsenko, Marco Picco, Paul Windey, 
                                  Geoffrey Harris, Enzo Marinari and Emil Martinec,
                     Quantum Field Theory and String Theory, Edited by L.Baulieu et             
                      al., Plenum Press, New York, 1995.            

                             





                      


                    
                               
\end{thebibliography}
\end{document}